\documentstyle[12pt]{article}

\newcommand{\EQ}{\begin{equation}}

\newcommand{\EN}{\end{equation}}

\newcommand{\ket}[1]{\left|#1\right\rangle}      

\newcommand{\bear}{\begin{eqnarray}}
\newcommand{\ear}{\end{eqnarray}}

\begin{document}

\topmargin 0pt
\oddsidemargin 5mm
\newcommand{\NP}[1]{Nucl.\ Phys.\ {\bf #1}}
\newcommand{\PL}[1]{Phys.\ Lett.\ {\bf #1}}
\newcommand{\NC}[1]{Nuovo Cimento {\bf #1}}
\newcommand{\CMP}[1]{Comm.\ Math.\ Phys.\ {\bf #1}}
\newcommand{\PR}[1]{Phys.\ Rev.\ {\bf #1}}
\newcommand{\PRL}[1]{Phys.\ Rev.\ Lett.\ {\bf #1}}
\newcommand{\MPL}[1]{Mod.\ Phys.\ Lett.\ {\bf #1}}
\newcommand{\JETP}[1]{Sov.\ Phys.\ JETP {\bf #1}}
\newcommand{\TMP}[1]{Teor.\ Mat.\ Fiz.\ {\bf #1}}
     
\renewcommand{\thefootnote}{\fnsymbol{footnote}}
     
\setcounter{page}{0}
\begin{titlepage}     
\begin{flushright}
UFSCAR-TH-22 
\end{flushright}
\vspace{0.5cm}
\begin{center}
\large{Algebraic Bethe ansatz for a class of coupled asymmetric six-vertex free-fermion
model
 } \\
\vspace{1cm}
\vspace{1cm}
{\large M.J. Martins} \\
\vspace{1cm}
\centerline{\em Departamento de F\'isica, Universidade Federal de S\~ao Carlos}
\centerline{\em Caixa Postal 676, 13565-905, S\~ao Carlos, Brazil}
\vspace{1.2cm}   
\end{center} 
\begin{abstract}
We present an algebraic Bethe ansatz for certain submanifolds of the bilayer vertex models 
proposed by Shiroishi and Wadati as coupled asymmetric six-vertex free-fermion models.
A peculiar feature of our formulation is the presence of a diagonal monodromy matrix element
that does not generate unwanted terms. The model contains two free-parameters entering
into the Bethe ansatz equations as a pure phase factor.
\end{abstract}
\vspace{.2cm}
\centerline{PACS numbers: 05.50.+q, 75.10.Hk,03.65.Fd }
\vspace{.2cm}
\centerline{Keywords: Integrable models, Algebraic Bethe ansatz}
\vspace{.2cm}
\centerline{September 2002}
\end{titlepage}

\renewcommand{\thefootnote}{\arabic{footnote}}

The six-vertex model satisfying the free-fermion condition possesses certain special properties
which are not common to general integrable vertex models of statistical mechanics \cite{WU}. 
This model provides a three-parameter family of commuting transfer matrices \cite{FE,BA} whose 
Boltzmann weights
satisfy generalized forms of the star-triangle relations \cite{SA,KO}. The later property
is the basis to generate
new exactly solvable vertex models by coupling in an appropriate way a pair of six-vertex
free-fermion models \cite{SA,KO}. 
This remarkable possibility has originally been formulated  by Shastry \cite{SA}
in the case of totally symmetric six-vertex free-fermion model. As a consequence of that
Shastry was able to rederive the
$R$-matrix of the covering vertex model for the one-dimensional Hubbard chain from a rather
promissing perspective \cite{SA}. 

More recently, Shiroishi and Wadati \cite{SW} extended such an approach to include the
asymmetric six-vertex free-fermion model which have resulted in new interacting bilayer vertex systems.
To our knowledge, however, the Bethe ansatz solution of such potentially interesting integrable
bilayer systems is still an open problem. 
The purpose  of this letter  is to make  a first step
toward solving this problem by formulating
the algebraic Bethe ansatz for what we expect to be the simplest branches of the
bilayer vertex models discovered by Shiroishi and Wadati \cite{SW}.  
These submanifolds  are interesting both because 
it does not contain the covering Hubbard model and due to an unusual 
feature concerning the
quantum inverse scattering method. It turns out that one of the diagonal elements of
the monodromy matrix 
does not produce
unwanted terms thanks to a remarkable annihilation property of the eigenvectors. 
The Bethe ansatz equations are of rational type with 
an additional phase-shift that encodes all possible free-parameters dependence. 

The building blocks of the bilayer vertex model constructed by Shiroishi
and Wadati are two trigonometric Felderhof's free-fermion
model \cite{FE} whose Lax operators are defined by
\EQ 
L_{{\cal{A}}i}^{(\alpha)}(\lambda,\gamma_i,\gamma_{i+1}) = 
\pmatrix{
\omega_a^{+}(\lambda,\gamma_i,\gamma_{i+1}) I_i + 
\omega_a^{-}(\lambda,\gamma_i,\gamma_{i+1}) \sigma^{z}_{\alpha,i}
&  \sigma_{\alpha,i}^{-}  \cr
\sigma_{\alpha,i}^{+}  &
\omega_b^{+}(\lambda,\gamma_i,\gamma_{i+1})I_i + 
\omega_b^{-}(\lambda,\gamma_i,\gamma_{i+1}) \sigma^{z}_{\alpha,i}
\cr} 
\nonumber \\
\EN
where $\sigma_{\alpha,i}^{\pm}$, 
$\sigma_{\alpha,i}^{z}$ are two sets $\alpha=\uparrow,\downarrow$ of commuting Pauli matrices and $I_i$
is the identity acting on the
site $i$ of a lattice of size $N$.  Here $\cal{A}$ denotes the auxiliary space,
$\lambda$ is the spectral parameter and $\gamma_k$ play the role
of color variables that are attached to each $k$-th site of the lattice.  The functions 
$\omega_a^{\pm}(\lambda,\gamma_i,\gamma_{j}) 
=\frac{a_{+}(\lambda,\gamma_i,\gamma_{j}) \pm
b_{+}(\lambda,\gamma_i,\gamma_{j}) }{2}$ and 
$\omega_b^{\pm}(\lambda,\gamma_i,\gamma_{j}) 
=\frac{b_{-}(\lambda,\gamma_i,\gamma_{j}) \pm
a_{-}(\lambda,\gamma_i,\gamma_{j}) }{2}$  where 
$a_{\pm}(\lambda,\gamma_i,\gamma_{j})$ 
and $b_{\pm}(\lambda,\gamma_i,\gamma_{j})$  are the elementary Boltzmann weights 
parameterized in terms of the spectral parameter $\lambda$ and
the color variables $\gamma_{k}$ by
\begin{equation}
a_{\pm}(\lambda,\gamma_i,\gamma_j)= \cosh(\gamma_i-\gamma_j) \cos(\lambda) \pm
\sinh(\gamma_i+\gamma_j) \sin(\lambda) 
\end{equation}
\begin{equation}
b_{\pm}(\lambda,\gamma_i,\gamma_j)= \cosh(\gamma_i+\gamma_j) \sin(\lambda) \mp
\sinh(\gamma_i-\gamma_j) \cos(\lambda) 
\end{equation}

The Lax operator of the bilayer vertex model is 
given by a suitable coupling of two Felderhof's free-fermion models and it can be 
written as follows 
\bear
{\cal{L}}_{{\cal{A}}i}(\lambda_i,\lambda_{i+1},\gamma_i,\gamma_{i+1}) & =& 
L_{{\cal{A}}i}^{(\uparrow)}(\lambda_i-\lambda_{i+1},\gamma_i,\gamma_{i+1})
L_{{\cal{A}}i}^{(\downarrow)}(\lambda_i-\lambda_{i+1},\gamma_i,\gamma_{i+1})
\nonumber \\
&& +f_1 
L_{{\cal{A}}i}^{(\uparrow)}(\lambda_i+\lambda_{i+1},\gamma_i,-\gamma_{i+1}) \sigma_{\uparrow,\cal{A}}^{z}
L_{{\cal{A}}i}^{(\downarrow)}(\lambda_{i}+\lambda_{i+1},\gamma_i,-\gamma_{i+1}) \sigma_{\downarrow,\cal{A}}^{z}
\nonumber \\
&& +f_2 [
L_{{\cal{A}}i}^{(\uparrow)}(\lambda_i+\lambda_{i+1},\gamma_i,-\gamma_{i+1}) \sigma_{\uparrow,\cal{A}}^{z}
L_{{\cal{A}}i}^{(\downarrow)}(\lambda_i-\lambda_{i+1},\gamma_i,\gamma_{i+1}) 
\nonumber \\
&& +L_{{\cal{A}}i}^{(\uparrow)}(\lambda_i-\lambda_{i+1},\gamma_i,\gamma_{i+1}) 
L_{{\cal{A}}i}^{(\downarrow)}(\lambda_i+\lambda_{i+1},\gamma_i,-\gamma_{i+1}) \sigma_{\downarrow,\cal{A}}^{z}]
\ear

The solution found by Shiroishi and Wadati \cite{SW} for the couplings
$f_{1,2}$ depends on two arbitrary constants denoted by 
$A$ and $B$ in their work. Here we are going to consider the simplest submanifolds as far as
the functional behavior of couplings is concerned. They occur when one of the constants is null because
the
other can be rescaled to unity without loosing generality. Even in this situation there exists
at least two free-parameters since the Lax operator (4) depends on both the difference and the
sum of the spectral parameters and color variables.     
It is convenient to
define the couplings of these submanifolds with the help of the relations
$g_{\pm}(\lambda,\gamma)=\frac{b_{\pm}(\lambda,\gamma,0)}{a_{\pm}(\lambda,\gamma,0)}$.  In terms
of these auxiliary functions they are given by
\begin{eqnarray}
f_1= \left \{ \begin{array}{ll}
               \displaystyle{
\frac{1}{4}
[\frac{g_{+}(\lambda_{i+1},\gamma_{i+1})}{g_{+}(\lambda_i,\gamma_i)} 
+\frac{g_{-}(\lambda_{i+1},\gamma_{i+1})}{g_{-}(\lambda_i,\gamma_i)} -2]}
& \mbox{ for B=0} \\
               \displaystyle{
\frac{1}{4}
[\frac{g_{+}(\lambda_{i},\gamma_{i})}{g_{+}(\lambda_{i+1},\gamma_{i+1})} 
+\frac{g_{-}(\lambda_{i},\gamma_{i})}{g_{-}(\lambda_{i+1},\gamma_{i+1})}-2]} 
& \mbox{ for A=0}
                \end{array} \right. 
\end{eqnarray}
and 
\begin{eqnarray}
f_2= \left \{ \begin{array}{ll}
               \displaystyle{
\frac{1}{4}
[\frac{g_{+}(\lambda_{i+1},\gamma_{i+1})}{g_{+}(\lambda_i,\gamma_i)} 
-\frac{g_{-}(\lambda_{i+1},\gamma_{i+1})}{g_{-}(\lambda_i,\gamma_i)} ]}
& \mbox{ for B=0} \\
               \displaystyle{
\frac{1}{4}
[\frac{g_{-}(\lambda_{i},\gamma_{i})}{g_{-}(\lambda_{i+1},\gamma_{i+1})} 
-\frac{g_{+}(\lambda_{i},\gamma_{i})}{g_{+}(\lambda_{i+1},\gamma_{i+1})}]} 
& \mbox{ for A=0}
                \end{array} \right. 
\end{eqnarray}

The fact that the 
$L$-operator (4) satisfies a colored version of the Yang-Baxter equation \cite{MU} implies 
that it is possible to define  a family of row-to-row
commuting transfer matrices $T(\lambda)$ as the trace of the monodromy matrix ${\cal{T}}(\lambda)$ 
on the auxiliary space of the
coupled six vertex free-fermion models. The 
monodromy matrix  then reads 
\begin{equation}
{\cal T}(\lambda) = {\cal L}_{ {\cal A} L}(\lambda,\lambda_0,\gamma,\gamma) 
{\cal L}_{ {\cal A} L-1}(\lambda,\lambda_0,\gamma,\gamma)  \cdots
{\cal L}_{ {\cal A} 1}(\lambda,\lambda_0,\gamma,\gamma) 
\end{equation}

These square lattice vertex models have two free parameters 
$\lambda_0$ and $\gamma$ and
their integrability is guaranteed by the following intertwining relation
\begin{equation}
R(\lambda,\mu) {\cal T}(\lambda) \otimes {\cal T}(\mu)=
{\cal T}(\mu) \otimes {\cal T}(\mu) R(\lambda,\mu)
\end{equation}
where the  $R$-matrix is  
$R(\lambda,\mu) = P_{12} {\cal L}_{12}(\lambda,\mu,\gamma,\gamma)$
and $P_{12}$ is the 
permutation operator.

We now turn to the algebraic 
Bethe ansatz  solution of these vertex models. The technicalities entering
the solution of the submanifolds $A=0$ and $B=0$ are very
similar and these details will be presented only for the former case. The 
main results, however,  such as  
the eigenvalues of the transfer matrices and corresponding Bethe equations will be given
for both of them.
An essential ingredient in this approach 
is the existence of a reference state in which
the monodromy matrix acts triangularly \cite{FA,KOP}. This helps us to identify 
the elements of the monodromy
matrix as particle creation and annihilation fields.  Here we denote this state by $\ket{0}$ and 
choose it as the standard ferromagnetic vacuum where all the spins are in the up eigenstate of
$\sigma_{\alpha,i}^{z}$. The action of the monodromy matrix in this state yields the relation
\EQ 
{\cal T}(\lambda) \ket{0} = 
\pmatrix{
[\omega_1(\lambda)]^{N}  &  \ddagger  & \ddagger & \ddagger  \cr
0  &  [\omega_2(\lambda)]^{N}   &  0 & \ddagger  \cr
0  &  0  &  [\omega_2(\lambda)]^{N}  & \ddagger  \cr
0  &  0  &  0  & [\omega_3(\lambda)]^{N}   \cr} \ket{0}
\EN
where the symbol $\ddagger$ denotes non-null values  and  
the functions $\omega_1(\lambda)$, 
$\omega_2(\lambda)$ and $\omega_3(\lambda)$ are given by
\EQ
\omega_1(\lambda)=a_{+}^2(\lambda-\lambda_0,\gamma,\gamma)
+a_{+}(\lambda+\lambda_0,\gamma,-\gamma)[f_1
a_{+}(\lambda+\lambda_0,\gamma,-\gamma)
+2f_2 a_{+}(\lambda-\lambda_0,\gamma,\gamma)
]
\EN
\bear
\omega_2(\lambda) & =& a_{+}(\lambda-\lambda_0,\gamma,\gamma)
b_{-}(\lambda-\lambda_0,\gamma,\gamma)
+f_1 a_{+}(\lambda+\lambda_0,\gamma,-\gamma)
b_{-}(\lambda+\lambda_0,\gamma,-\gamma)
\nonumber \\
&&
f_2 [a_{+}(\lambda+\lambda_0,\gamma,-\gamma)
b_{-}(\lambda-\lambda_0,\gamma,\gamma)
+a_{+}(\lambda-\lambda_0,\gamma,\gamma)
b_{-}(\lambda+\lambda_0,\gamma,-\gamma)]
\nonumber \\
\ear
\EQ
\omega_3(\lambda)=b_{-}^2(\lambda-\lambda_0,\gamma,\gamma)
+b_{-}(\lambda+\lambda_0,\gamma,-\gamma)[f_1
b_{-}(\lambda+\lambda_0,\gamma,-\gamma)
+2f_2 b_{-}(\lambda-\lambda_0,\gamma,\gamma)
]
\EN

The next step is to seek for a suitable representation of the monodromy 
matrix.  Previous experience in dealing with  the quantum inverse scattering method for 
genuine four dimensional
representation \cite{MAR} 
suggests us to take the following form
\EQ
{\cal T}(\lambda) = 
\pmatrix{
B(\lambda)       &   \vec{B}(\lambda)   &   F(\lambda)   \cr
\vec{C}(\lambda)  &  \hat{A}(\lambda)   &  \vec{B^{*}}(\lambda)   \cr
C(\lambda)  & \vec{C^{*}}(\lambda)  &  D(\lambda)  \cr}_{4 \times 4}
\EN
where as in ref.\cite{MAR} $\vec{B}(\lambda)$,  
$\vec{B^{*}}(\lambda)$, 
$\vec{C}(\lambda)$ and
$\vec{C^{*}}(\lambda)$ are two dimensional vectors, the field
$ \hat{A}(\lambda)  $ is a $ 2 \times 2$ matrix and the remaining operators are scalars.

Comparison between the above representation and the triangular property (9) reveals that
$\vec{B}(\lambda)$,  
$\vec{B^{*}}(\lambda)$ and $F(\lambda)$ play the role of creation operators with respect to our choice of
reference state. This by no means is an indication that all of them are needed to generate the 
eigenstates of the transfer matrix. To make further progress we have to investigate the commutation relations
between the monodromy operators from the Yang-Baxter algebra (7). In order to present the results
as general as possible we find convenient
to represent the $R$-matrix by $R(\lambda,\mu)= \displaystyle{\sum_{a,b,c,d}^{4}} R_{ab}^{cd}(\lambda,\mu) e_{ac} \otimes 
e_{bd} $ where $e_{ab}$ are the Weyl matrices.  
We first look at the commutation rules
between the creation operator
$\vec{B}(\lambda)$ with itself and with the diagonal fields and the results are 
\EQ
\vec{B}(\lambda) \otimes \vec{B}(\mu) = \frac{R_{22}^{22}(\lambda,\mu)}{R_{11}^{11}(\lambda,\mu)}
[ \vec{B}(\mu) \otimes \vec{B}(\lambda) ] .\hat{r}(\lambda,\mu)
\EN
\EQ
B(\lambda)\vec{B}(\mu) = 
\frac{R_{11}^{11}(\mu,\lambda)}{R_{12}^{21}(\lambda,\mu)} \vec{B}(\mu)B(\lambda) - 
\frac{R_{21}^{21}(\mu,\lambda)}{R_{12}^{21}(\lambda,\mu)} \vec{B}(\lambda)B(\mu)
\EN
\EQ
\hat{A}(\lambda) \otimes \vec{B}(\mu) = 
\frac{R_{22}^{22}(\lambda,\mu)}{R_{12}^{21}(\lambda,\mu)}[\vec{B}(\mu) \otimes 
\hat{A}(\lambda) ]. \hat{r}(\lambda,\mu)
-\frac{R_{12}^{12}(\lambda,\mu)}{R_{12}^{21}(\lambda,\mu)} \vec{B}(\lambda) \otimes \hat{A}(\mu)   
\EN
\EQ
D(\lambda)\vec{B}(\mu)  = 
\frac{R_{24}^{42}(\lambda,\mu)}{R_{13}^{31}(\lambda,\mu)} \vec{B}(\mu)D(\lambda) 
+ \frac{R_{42}^{42}(\lambda,\mu)}{R_{13}^{31}(\lambda,\mu)} F(u)\vec{C^{*}}(\lambda)
 \nonumber \\ 
 - \frac{R_{14}^{14}(\lambda,\mu)}{R_{13}^{31}(\lambda,\mu)} F(\lambda)\vec{C^{*}}(\mu)
\EN

The auxiliary  matrix 
$\hat{r}(\lambda,\mu)$ is given by
\EQ
\hat{r}(\lambda,\mu) = 
\pmatrix{
1  &0  &0  &0  \cr
0  &a(\lambda,\mu)  &b(\lambda,\mu)  &0  \cr
0  &b(\lambda,\mu)  &a(\lambda,\mu)  &0  \cr
0  &0  &0  &1  \cr}
\EN
whose 
non-trivial matrix elements
$a(\lambda,\mu)$ and $b(\lambda,\mu)$ are
\EQ
a(\lambda,\mu)= \frac{\frac{2}{\cosh(2\gamma)}}{\frac{2}{\cosh(2\gamma)}+\cos(2\lambda)-\cos(2\mu)},~~
b(\lambda,\mu)= -\frac{\cos(2\lambda)-\cos(2\mu)}{\frac{2}{\cosh(2\gamma)}+\cos(2\lambda)-\cos(2\mu)}
\EN

From these results we first recognize  that the matrix 
$\hat{r}(\lambda,\mu)$ is related to the one of the isotropic six-vertex model via reparametrization of the weights
in terms of a new variable $\tilde{\lambda}=\cos(2\lambda)$. Next we observe from the commutation rules (14-16) that
two creation fields 
$\vec{B}(\lambda)$ can not generate other creations fields and also that
all the unwanted terms coming from the diagonal fields $B(\lambda)$ and 
$\hat{A}(\lambda) $ remain in the space of the states of the operator  
$\vec{B}(\lambda)$. It is notable that such commutation relations are in the same form of that
appearing in the nested Bethe ansatz solutions of multistate generalizations of the six-vertex
model \cite{RE,DE}.
A relevant difference, however, is the
presence of the diagonal field $D(\lambda)$  which requires further
elaboration. As far as the one-particle state is concerned, the commutation relation (16) shows
that this operator generates unwanted terms that are trivially canceled out due to the
annihilation property
$ \vec{C^{*}}(\lambda)  \ket{0}=0$. To show that this is indeed a general situation we need
the help of an extra commutation rule, namely
\EQ
\vec{C^{*}}(\lambda) \otimes \vec{B}(\mu) = 
\frac{R_{22}^{22}(\lambda,\mu)}{R_{13}^{31}(\lambda,\mu)}
\hat{r}(\lambda,\mu).
\vec{B}(\mu) \otimes 
\vec{C^{*}}(\lambda) 
-\frac{R_{14}^{14}(\lambda,\mu)}{R_{13}^{31}(\lambda,\mu)} \vec{B}(\lambda) \otimes \vec{C^{*}}(\mu)   
\EN

Now by iterating this relation on the tensor product of several 
$\vec{B}(\lambda_i) $ we are able to derive the following remarkable annihilation  property
\EQ
\vec{C^{*}}(\lambda) \otimes \vec{B}(\lambda_1) \otimes
\ldots \otimes \vec{B}(\lambda_n) \ket{0} =0
\EN
which together with the commutation rule (17) implies that the field $D(\lambda)$ does not
produce unwanted terms in the space of states  generated by the creation fields
$\vec{B}(\lambda)$. 

These observations strongly suggest us to take as the eigenvectors of the transfer matrix 
$T(\lambda)$ the following linear combination
\EQ
\ket{\Phi(\lambda_{1}, \ldots ,\lambda_{n})} =  
\vec{B}(\lambda_1) \otimes
\vec{B}(\lambda_2) \otimes
\ldots \otimes \vec{B}(\lambda_n) 
.\vec{\cal{F}} \ket{0}
\EN
where the vector
$\vec{\cal{F}}$ depends on the parameter $\lambda_i$ and its components
will be denoted by
${\cal{F}}^{a_{1} \cdots a_{n}} $ where each $a_i=1,2$. 

At this point we have gathered the basic tools to determine the eigenvalues 
$\Lambda(\lambda,\{\lambda_i\})$ of the transfer matrix defined
by
\EQ
[B(\lambda)+\sum_{a=1}^{2}A_{aa}(\lambda)+D(\lambda)]  
\ket{\Phi(\lambda_{1}, \ldots ,\lambda_{n})} = 
\Lambda(\lambda,\{ \lambda_i \}) 
\ket{\Phi(\lambda_{1}, \ldots ,\lambda_{n})} 
\EN

In order to do that we use the commutation rules (15-17) and the property (21) to commute the diagonal
operator through all the 
$\vec{B}(\lambda_i) $ fields until they hit the reference state.  Their values acting on the reference state
can be read out from Eq.(9). 
Omitting details of the calculations
our final result for the eigenvalues of the transfer matrix is
\EQ
\Lambda(\lambda,\{\lambda_{j}\}) = 
[\omega_1(\lambda)]^ N
\prod_{j=1}^{n} \frac{R_{11}^{11}(\lambda_{j},\lambda)}{R_{12}^{21}(\lambda_{j},\lambda)} + 
[\omega_3(\lambda)]^N 
\prod_{j=1}^{n} \frac{R_{24}^{42}(\lambda,\lambda_{j})}{R_{13}^{31}(\lambda,\lambda_{j})}  
\nonumber\\
+[\omega_2(\lambda)]^N \prod_{j=1}^{n} \frac{R_{22}^{22}(\lambda,\lambda_{j})}{R_{12}^{21}(\lambda,\lambda_{j})} 
\Lambda^{(1)}(\lambda,\{\lambda_{l}\}) 
\EN
provided that the rapidities  satisfy the Bethe 
equations
\EQ
\left[ \frac{\omega_1(\lambda_{i})}{\omega_2(\lambda_{i})} \right]^{L} = (-1)^{n-1} 
\Lambda^{(1)}(\lambda=\lambda_{i},\{\lambda_{j}\}), ~~ i=1, \ldots, n
\EN

The function
$\Lambda^{(1)}(\lambda,\{ \lambda_l \})$ refers to the eigenvalue of an auxiliary problem related
to the condition that the vector $\vec{\cal{F}}$ be an eigenvector of the transfer matrix whose
Boltzmann weights are the $R$-matrices (18), namely
\EQ
\hat{r}_{b_{1}d_{1}}^{c_{1}a_{1}}(\lambda,\lambda_{1})
\hat{r}_{b_{2}c_{2}}^{d_{1}a_{2}}(\lambda,\lambda_{2}) \ldots
\hat{r}_{b_{n}c_{1}}^{d_{n-1}a_{n}}(\lambda,\lambda_{n})
{\cal{F}}^{a_{n} \cdots a_{1}}    =
\Lambda^{(1)}(\lambda,\{\lambda_{j}\})
{\cal{F}}^{b_{n} \cdots b_{1}}    
\EN

This problem is solved by another Bethe ansatz which gives the well known results \cite{FA,KOP}
for the inhomogeneous rational six-vertex model,
\EQ
\Lambda^{(1)}(\lambda,\{\lambda_{i}\},\{ \mu_j \}) =
\prod_{j=1}^{m} \frac{1}{b(\mu_j,\lambda)} +
\prod_{i=1}^{n} b(\lambda,\lambda_i) \prod_{j=1}^{m} \frac{1}{b(\lambda,\mu_j)}
\EN
where the auxiliary variables $ \{ \mu_j \} $ are subject to the following set of constraints
\EQ
\prod_{i=1}^{n} b(\mu_j,\lambda_i) = - \prod_{k=1}^{m} \frac{b(\mu_j,\mu_k)}{b(\mu_k,\mu_j)},~~
j=1, \cdots, m
\EN

Now we first insert the nested eigenvalues (27) into Eq.(24). By taking into account the explicit expression of the
$R$-matrix elements after carrying out cumbersome simplifications we find that
\bear
\Lambda(\lambda,\{ \lambda_{j} \},\{ \mu_{l} \}) & = &
[\omega_1(\lambda)]^L \prod_{j=1}^{n} z(\lambda,\gamma)
\left [ \frac{U-\cos(2\lambda_i)+\cos(2\lambda)}{\cos(2\lambda_i) -\cos(2\lambda)} \right ]
+[\omega_3(\lambda)]^L \prod_{j=1}^{n}  z(\lambda,\gamma)
 \nonumber \\
&& +[\omega_2(\lambda)]^{L} \left \{
\prod_{j=1}^{n} z(\lambda,\gamma)
\left [ \frac{U+\cos(2\lambda)-\cos(2\lambda_i)}
{\cos(2\lambda)-\cos(2\lambda_i)} \right ]
\prod_{l=1}^{m} \frac{ U+\cos(2\mu_l)-\cos(2\lambda)}{\cos(2\lambda)-\cos(2\mu_l)}
\right.
\nonumber\\ 
&& \left. +\prod_{j=1}^{n}  -z(\lambda,\gamma)
\prod_{l=1}^{m}
 \left [ \frac{U+\cos(2\lambda) -\cos(2\mu_l)}{\cos(2\mu_l)-\cos(2\lambda)}
 \right ]
\right \}
\ear
where the function $z(\lambda,\gamma)$ and the parameter $U$ are given by
\begin{eqnarray}
z(\lambda,\gamma)= \left \{ \begin{array}{ll}
               \displaystyle{\frac{b_{+}(\lambda,\gamma,0)}{a_{-}(\lambda,\gamma,0)}} & \mbox{ for B=0} \\
               \displaystyle{-\frac{a_{+}(\lambda,\gamma,0)}{b_{-}(\lambda,\gamma,0)}} & \mbox{ for A=0}
                \end{array} \right. ~~~{\rm and}~~~
U= \left \{ \begin{array}{ll}
     \displaystyle{\frac{2}{\cosh(2\gamma)}} & \mbox{ for B=0} \\
     \displaystyle{-\frac{2}{\cosh(2\gamma)}} & \mbox{ for A=0} 
\end{array} \right.
\end{eqnarray}

Proceeding similarly for Eqs.(25,28) we obtain the corresponding nested Bethe ansatz equations
\bear
\left [\frac{U -\cos(2\lambda_i) + \cos(2\lambda_0)}
{-\cos(2\lambda_i) + \cos(2\lambda_0)}
\right ]^N  [-z(\lambda_0,\gamma_0)]^{N}& = & (-1)^{n-1}
\prod_{l=1}^{m}  \frac{U+\cos(2\mu_l) -\cos(2\lambda_i)}{\cos(2\lambda_i)-\cos(2\mu_l)},
~j=1, \ldots, n
\nonumber\\
\prod_{j=1}^{n}  \frac{\cos(2\lambda_j)-\cos(2\mu_l)}{U+\cos(2\mu_l)-\cos(2\lambda_j)} & =&
-\prod_{k=1}^{m} \frac{\cos(2\mu_l)-\cos(2\mu_k)-U}{\cos(2\mu_l)-\cos(2\mu_k)+U},~l=1, \ldots, m 
\nonumber \\
\ear

From the result (31) we observe that it is possible 
to cast the Bethe equations in a simpler
form  if we define the following new rapidities 
\EQ
\frac{\cos(2\lambda_i)}{U} = \tilde{\lambda_i} +\frac{1}{2} +\frac{\cos(2\lambda_0)}{U},~~
\frac{\cos(2\mu_i)}{U} = \tilde{\mu_i} + \frac{\cos(2\lambda_0)}{U}
\EN
leading us to the expressions
\bear
\left [\frac{\tilde{\lambda}_i - \frac{1}{2}}
{\tilde{\lambda}_i + \frac{1}{2}}
\right ]^N  [-z(\lambda_0,\gamma_0)]^{N}& = & (-1)^{n+m-1}
\prod_{l=1}^{m}  \frac{\tilde{\lambda}_i -\tilde{\mu}_l -\frac{1}{2}}
{\tilde{\lambda}_i -\tilde{\mu}_l +\frac{1}{2}}
~j=1, \ldots, n
\nonumber\\
(-1)^{n} \prod_{j=1}^{n}  \frac{\tilde{\mu}_l-\tilde{\lambda}_j -\frac{1}{2}}{\tilde{\mu}_l -\tilde{\lambda}_j+\frac{1}{2}} & =&
-\prod_{k=1}^{m}  \frac{\tilde{\mu}_l-\tilde{\mu}_k -1}{\tilde{\mu}_l -\tilde{\mu}_k+1} 
,~l=1, \ldots, m 
\ear

In terms of these variables the dependence of the Bethe equations on the free-parameters becomes
restricted as a fixed phase factor. Next we observe that their rational form resemble that of
the supersymmetric $t$-$J$ model \cite{TJ1,TJ2} 
with the important difference 
that in our case we have a bigger space of states such that
$m \leq N$ rather than 
$m \leq N/2$. As far as the Hilbert space is concerned one would
be tempted to compare these equations with that of the supersymmetric $U$ model \cite{LI}. 
However, this demands that the  continuous parameter governing the four dimensional representation
of the superalgebra $spl(2|1)$ reaches the value related to atypical representations which is 
not permitted in this model \cite{MAS,LI1}. These remarks are strong 
evidences that the class of models solved
here lies in between these two supersymmetric systems motivating subsequent investigations. 
One of them  consists in the computation of physical properties of the bilayer vertex model such as the 
ground
state behavior dependence on the free-parameters. Another interesting point raised here is
the possible connection  between this system and 
four dimensional atypical $spl(2|1)$ representations.  It seems also worthwhile to generalize our
results to include the manifold 
discovered by Shiroishi and Wadati for arbitrary values of the constants $A$ and $B$. It is plausible
to believe that this will 
lead us to Bethe equations interpolating between that of the $t$-$J$ and Hubbard like models.
In this situation a 
crucial point is to unveil the corresponding six-vertex hidden symmetry which has eluded us so far.

\section*{Acknowledgements} 
This work has been partially supported
by the Brazilian research Agencies CNPq and Fapesp.

\end{document}